\documentclass[aps,showpacs,nofootinbib,12pt]{revtex4}
\usepackage{amsmath}
\usepackage{amsfonts}
\usepackage{amssymb}
\usepackage{color}

\def\bc{\begin{center}}
\def\nno{\nonumber}
\def\ec{\end{center}}
\def\be{\begin{eqnarray}}
\def\ee{\end{eqnarray}}

\definecolor{dyellow}{rgb}{1.,0.8,.0}
\definecolor{myblue}{rgb}{.1,.1,.7}
\definecolor{dcyan}{rgb}{.0,.6,.6}
\definecolor{dmagenta}{rgb}{0.6,0.0,0.6}
\definecolor{brown}{rgb}{0.6,0.2,0.}
\definecolor{darkblue}{rgb}{.0,.0,0.5}
\definecolor{darkred}{rgb}{0.75,0.0,0.0}
\definecolor{orange}{rgb}{1.,.6,.0}
\definecolor{dorange}{rgb}{0.8,.4,.0}
\definecolor{darkgreen}{rgb}{0.0,0.6,0.0}
\definecolor{purple}{rgb}{.4,.0,.4}
\definecolor{lightgrey}{rgb}{0.7, 0.7, 0.7}
\definecolor{grey}{rgb}{0.4, 0.4, 0.4}

\def\La{\Lambda}

\def\ga{\gamma}

\def\ka{\kappa}
\def\la{\lambda}

\def\si{\sigma}

\def\pa{\partial}

\newcommand\btd{\raise 2pt
\hbox{$\hat\bigtriangledown$}\hskip 1.5pt}
\newcommand\bt{\raise 2pt
\hbox{$\bigtriangledown$}\hskip 1.5pt}

\newcommand{\omits}[1]{}

\def\NPB{{Nucl. Phys.}~{\bf B}}
\def\PRD{{Phys. Rev.}~{\bf D}}
\def\PRL{{Phys. Rev. Lett. }}

\def\PLB{{Phys. Lett.}~{\bf B}}

\def\CQG{{Class. Quant. Grav. }}

\def\CTP{{Commun. Theor. Phys. }}

\def\IJMPD{{Int. J. Mod. Phys.}~{\bf D}}
\def\IJMPA{{Int. J. Mod. Phys.}~{\bf A}}

\begin{document}

\title{Gravitational Anomaly and Hawking Radiation of\\ Brane World Black Holes}

\vskip 2cm

\author{Chao-Guang Huang$^{a,c}$\footnote{Email: huangcg@ihep.ac.cn}, Jia-Rui Sun$^{a,d}$\footnote{Email: sun@ihep.ac.cn}, Xiaoning Wu$^{b,c}$\footnote{Email: wuxn@amss.ac.cn},
and Hai-Qing Zhang$^{a,d}$
} \affiliation{\footnotesize $^a$Institute of High Energy Physics,
Chinese Academy of Sciences, P.O.Box 918(4), Beijing, 100049,
China} \affiliation{\footnotesize $^b$Institute of Mathematics,
Academy of Mathematics and Systems Science, Chinese Academy of
Sciences, P.O.Box 2734, Beijing, 100080, China}
\affiliation{\footnotesize $^c$Kavli Institute for Theoretical
Physics China  at the Chinese Academy of Sciences (KITPC-CAS),
P.O.Box 2732, Beijing, 100080, China} \affiliation{\footnotesize
$^d$Graduate School of Chinese Academy of Sciences, Beijing,
100049, China.}

\bigskip 

\date{October, 2007}

\newpage

\begin{abstract}
We apply Wilczek and his collaborators' anomaly cancellation
approach to the 3-dimensional Schwarzschild- and BTZ-like brane
world black holes induced by the generalized $\mathcal{C}$ metrics
in the Randall-Sundrum scenario. Based on the fact that the horizon
of brane world black hole will extend into the bulk spacetime, we do
the calculation from the bulk generalized $\mathcal{C}$ metrics side
and show that this approach also reproduces the correct Hawking
radiation for these brane world black holes. Besides, since this
approach does not involve the dynamical equation, it also shows that
the Hawking radiation is only a kinematic effect.
\end{abstract}


\maketitle


\section{Introduction}
It is often instructive to study a physical phenomenon from
different point of views.  The Hawking radiation is one of the
famous examples.  After Hawking's original derivation
\cite{Hawking}, there emerges many other approaches to recover this
effect (e.g.,\cite{unruh, dr, pw, cf}).  For example, Christensen
and Fulling use the trace anomaly to study the Hawking effect
\cite{cf}.  They consider a conformal scalar field in a
Schwarzschild black hole background and find that the anomalous
trace of the renormalized stress tensor of the scalar field can
reproduce the Hawking radiation under the requirement that the
renormalized stress tensor is conserved covariantly. Recently,
Wilczek and his collaborators propose a new approach to obtain the
Hawking radiation from the viewpoint of gravitational and gauge
anomalies \cite{rw, iuw1, iuw2}. Instead of the trace anomaly, they
concern the covariant anomaly and gauge anomaly in a reduced
2-dimensional spacetime, which falls into the class of quantum
fields in $4k+2$ ($k\in Z$) dimensional spacetime, studied by
Alvarez-Gaume and Witten \cite{witten}. Wilczek {\it et al} consider
a scalar field in a stationary black hole background and show that
Hawking radiation can be determined by the anomaly cancellation and
regularity condition at horizon. Subsequently, this method have been
extended into various black hole backgrounds \cite{ms, vd, se, xc,
jw, jwc, xlz, sk, drv, ch, mm, bk} and black hole-like background
\cite{ks}. All of these results indicate that the gravitational
anomaly has an intrinsic connection with the Hawking radiation and
that this method could be applied to the more general situations.

The purpose of the present letter is to check whether the anomaly
cancellation approach can be applied to the brane world black holes
induced by the generalized $\mathcal{C}$ metrics in the
Randall-Sundrum (RS) scenario \cite{RS}. We mainly focus on two
cases: 3-dimensional Schwarzschild- and BTZ-like black holes on RS
2-branes. Both of them are obtained from the generalized
$\mathcal{C}$ metrics. Like their 4-dimensional counterparts, these
brane world black holes also have Hawking radiation, and their
thermodynamics have been studied in \cite{ehm1}. So it is probably
that the anomaly cancellation approach also applies to these
situations. Generally speaking, one may directly use this method for
these brane world black holes from the brane world side. However, we
would like to do it in another way, i.e., from the bulk side. The
reason is that various studies show that the horizons of the brane
world black holes will extend into the extra dimension, but not
merely on the brane, and these brane world black holes can be
described as black strings in the bulk spacetime \cite{Chr, ehm1,
kanti, cfm, aliev}. For the two cases we consider here, the bulk is
described by the generalized ${\cal C}$ metric. It also contains the
black hole horizon, which is just the extension of the black hole
horizon on the brane \cite{emparan}. So one can apply the anomaly
cancellation method to the generalized ${\cal C}$ metric, instead of
directly applying it to the brane world black holes. For
convenience, we only consider the gravitational anomaly and obtain
the correct Hawking radiation for these brane world black holes.
Besides, since this method does not involve the dynamical equation,
i.e., the Einstein equation, it also shows that the Hawking
radiation is only a kinematic effect, this is in accord with the
studies about the acoustic black holes and the uniformly accelerated
reference frames \cite{unruh1, vi1, vi2, cg, Sun}.

The letter is organized in the following way.  In the next
section, we will briefly review the anomaly cancellation approach
proposed by Wilczek and his collaborators. In sections III and IV,
we will study the Hawking radiation of the 3-dimensional
Schwarzschild- and BTZ-like brane world black holes by applying
this approach to the generalized $\mathcal{C}$ metrics without and
with a rotation parameter, respectively. Then we will come to the
conclusions and discussions.


\section{A Brief Review of the Method}

The effective action of a massless scalar field in a
$d$-dimensional static, spherically symmetric black hole
background can be reduced to a 2-dimensional one for an infinite
collection of 2-dimensional scalar fields in the near horizon
region by using the tortoise coordinate \cite{rw}. The horizon
requires a boundary condition that the outgoing modes vanish near
it, thus the 2-dimensional effective fields become chiral in this
region and then the gravitational anomaly will appear as
\cite{witten, bert1, bert2}
\be \label{anomaly} \nabla_\mu T^\mu_\nu=
\frac{1}{96\pi\sqrt{-g}}\epsilon^{\beta \delta}\partial_\delta
\partial_\alpha\Gamma^\alpha_{\nu \beta},
\ee
where $\Gamma^\alpha_{\nu \beta}$ is the Christoffel connection of
the effective 2-dimensional metric
\be \label{2d metric} ds^2=-f(r)dt^2+f(r)^{-1}dr^2. \ee
Its horizon is located at $r=r_H^{}$, i.e., $f(r_H^{})=0$, and its
surface gravity is $\ka=|f'(r_H^{})|/2$. For Eq.(\ref{2d metric}),
the anomaly is purely timelike and can be rewritten as
\be \label{anomaly1} \nabla_\mu T^\mu_\nu \equiv A_\nu \equiv
\frac{1}{\sqrt{-g}}\partial_\mu N^\mu_\nu. \ee
Eq.(\ref{anomaly}) and Eq.(\ref{anomaly1}) give
$N^\mu_\nu=\frac{1}{96\pi}\epsilon^{\mu \lambda}\partial_\alpha
\Gamma^\alpha_{\lambda \nu}$, its components are
\be \label{N}
N^t_t&=&N^r_r=0,\nno \\
N^r_t&=&\frac{1}{192\pi}(f^{\prime 2}+f^{\prime\prime}f),\nno \\
N^t_r&=&\frac{-1}{192\pi}\frac{(f^{\prime
2}-f^{\prime\prime}f)}{f^2}. \ee

The anomaly appears when the effective action of the field at
quantum level fails to be diffeomorphism invariant, i.e.,
$\delta_\xi W=-\int{d^dx \sqrt{-g}\xi^\nu\nabla_\mu T^\mu_\nu}\neq
0$, where $\xi^a$ is an arbitrary vector field which generates the
transformation of action and $T^\mu_\nu$ is the stress tensor. If
the full quantum field theory is still required to be generally
covariant, namely $\delta_\xi W= 0$, then the flux of each
outgoing partial wave, which is added to cancel the anomaly, just
reproduces the Hawking radiation.

In detail, by use of two step functions $\Theta_+=\Theta(r - r_H
-\varepsilon)$ and $\Theta_-=1-\Theta_+ $, the variation of the
effective action can be written as
\be \label{delta W} %
-\delta_\xi W &=&\int{d^2x \sqrt{-g}\xi^\nu\nabla_\mu
\left\{ {T_\chi}^\mu_\nu \Theta_{-}  +{T_o}^\mu_\nu\Theta_{+}  \right\}}  \nno\\
&=&\int{d^2x\left\{\xi^t[ \partial_r (N^r_t\Theta_{-})
    + \left({T_o}^r_t-{T_\chi}^r_t+N^r_t\right)\partial_r\Theta_{+}
  ] + \xi^r [
    \left({T_o}^r_r-{T_\chi}^r_r\right)\partial_r\Theta_{+}
     ] \right\}},
\ee
where ${T_\chi}^\mu_\nu$ and ${T_o}^\mu_\nu$ are the stress tensor
in the regions $r_H^{}<r<r_H^{}+\varepsilon$ and
$r>r_H^{}+\varepsilon$, respectively. The integration of
Eq.(\ref{anomaly1}) gives rise to
\be \label{T}%
T^t_t&=& -(K+Q)/f-B(r)/f-I(r)/f+T^\alpha_\alpha(r),\nno\\
T^r_r&=&  (K+Q)/f+B(r)/f+I(r)/f,\nno \\
T^r_t&=& -K + C(r)  = -f^2 T^t_r, \ee
where $K$ and $Q$ are constant, while
$B(r)=\int_{r_H}^r{f(x)A_r(x)dx}$,
$I(r)=\frac{1}{2}\int_{r_H}^r{T^\alpha_\alpha(x)f'(x)dx}$ and
$C(r)=\int_{r_H}^r{A_t(x)dx}$. Substituting Eq.(\ref{T}) into
Eq.(\ref{delta W}), one gets \omits{and expanding it near the
black hole horizon up to the linear term of $\varepsilon$,}
\be \label{delta W1} %
-\delta_\xi W &=& \int d^2x\xi^t \left \{\partial_r
(N^r_t\Theta_{-})+
 \left[ N^r_t +K_\chi-K_o \right] 
\delta(r-r_H-\varepsilon)\right \}\nonumber\\
 && +\int d^2x\xi^r \left\{
   \frac{K_o+Q_o-K_\chi-Q_\chi }{f} \delta(r-r_H-\varepsilon)  \right\}. %
\ee
The requirement of $\displaystyle{\lim_{\varepsilon\to
0}}\delta_\xi W=0$ for any $\xi$ demands
\be \label{KQ}
K_o&=&K_\chi+\Phi,\nno\\
Q_o&=&Q_\chi-\Phi, \ee
where
\be \label{flux} \Phi=N^r_t|_{r_H^{}}=\frac{\kappa^2}{48\pi} \ee
is just the flux which is needed to cancel the gravitational
anomaly. Note that a beam of massless blackbody radiation moving
in the radial direction at temperature $T$ has the form
$\Phi=\frac{\pi}{12}T^2$, thus Eq.(\ref{flux}) just reproduce the
Hawking radiation.

For the scalar field in a charged or rotational black hole,
\cite{iuw1, iuw2, ms}, e.g., the Reissner-Nordstr\"{o}m (RN) black
hole or the Kerr-Newman black
hole, 
there will appear gauge anomaly in addition to the gravitational
anomaly. Consequently, these gauge anomalies will contribute to
the black body radiation (Hawking radiation) spectrum
(characterized by the distribution of particles $I^{(\pm)}$ at
given energy and chemical potentials) of the black hole as the
corresponding chemical potentials
\be \label{bbs} I^{(\pm)}=\frac{1}{e^{\beta (E\pm
\sum_i\mu_i)}+1}, \ee
where $\beta$ is the inverse Hawking temperature, $E$ is the
energy of the particle and $\mu_i$ are the chemical potentials.
Like the treatment in \cite{rw, iuw1, iuw2}, we only consider
Fermi distribution.


\section{Hawking Radiation of Schwarzschild-like black hole on 2-brane}
It is well known that the generalized $\mathcal{C}$ metrics are a
large class of degenerate vacuum (electro-vacuum) solutions for
the Einstein (Einstein-Maxwell) equations with or without a
cosmological constant, which may describe accelerating black holes
in spacetime \cite{cmetric, cbook, ehm1}. One of its subclasses is
known as $AdS$ $\mathcal{C}$ metric. Its line element can be
written as
\be \label{adscm} ds^2 &=& \frac{1}{(x-y)^2}[-H(y)dt^2
+\frac{dy^2}{H(y)}
+\frac{dx^2}{F(x)}+F(x)d\si^2] %
\ee
with
\be \label{hx}
 H(y)&=&-(\ga+\La/6) -2ly+ky^2-2My^3,\nno\\
F(x)&=&\ga-\La/6+2lx-kx^2+2Mx^3, \ee
where $k=1, 0, -1$, related to the topology of the horizon, $M$ is
the mass parameter of the black hole, $l$ is the NUT parameter
which is usually regarded as unphysical and can be chosen to be
zero, $\La (< 0)$ is the cosmological constant and $\ga$ is the
acceleration parameter. In order to keep the Lorentz signature
$(-,+,+,+)$ in metric (\ref{adscm}), one needs $F(x)\geq 0$.
Generally, this spacetime has one or more event horizons, which
are located at
\be \label{h} H(y)&=&-(\ga+\La/6) -2ly+ky^2-2My^3=0. \ee

To construct the 3-dimensional Schwarzchild-like brane world black
hole from the $AdS$ $\mathcal{C}$ metric, one just needs to set
$2\ga = -\La/3 =1$, $l=0$ and $k=1$ in Eq.(\ref{adscm}) firstly,
which result in
\be \label{hx1}
 H(y)&&=y^2-2My^3,\nno\\
X(x)&&=1-x^2+2Mx^3,%
\ee
and then make the coordinate transformations $z=1-x/y$ and
$\rho=\sqrt{1-x^2}/y$. After that, the 3-dimensional induced
metric on the RS 2-brane at $x=0$ is just a 3-dimensional
Schwarzschild-like brane world black hole
\be \label{sbh}
ds^2=-(1-\frac{2M}{\rho})dt^2+(1-\frac{2M}{\rho})^{-1}d\rho^2+\rho^2d\si^2.
\ee
As has been studied in \cite{ehm1}, its horizon on the 2-brane is
located at $\rho=\rho_h=2M$, and the Hawking temperature of the
horizon is $T_h=|f'(\rho)|/(4\pi)|_{\rho_{h}}=1/(8\pi M)$, which
is of the same form as that of the 4-dimensional Schwarzschild
black hole, and, this horizon will extend into the extra dimension
$x$, but not merely on the brane, due to the bulk/brane
interaction. Namely, the horizon of the 3-dimensional brane world
black hole is a higher dimensional object with spatial dimension
three. Notice that this brane world black hole is a static one.
Thus its horizon on the brane is a minimal surface. Besides, one
can show that the bulk extension of this brane world black hole
horizon is also a minimal surface, with the horizon on the brane
as its boundary \cite{emparan}. Furthermore, it can also be shown
that such a minimal surface is just the bulk black hole horizon in
$AdS$ $\mathcal{C}$ spacetime.
Consequently, it can be proved  $T_H=T_h$, where $T_H$ is the
Hawking temperature of the bulk black hole horiozn. Therefore, one
can apply the anomaly cancellation approach to the bulk black hole
horizon of the $AdS$ $\mathcal{C}$ metric, instead of directly
applying it to the 3-dimensional Schwarzschild-like brane world
black hole itself.

To do this, consider a massless scalar field $\phi$ in the $AdS$ $\mathcal{C}$ spacetime with the classical action 
\be \label{action} S =&&-\frac{1}{2}\int
d^4x\sqrt{-g}\phi\Box\phi\nno\\
=&&-\frac{1}{2}\int d^4x\phi[-\frac{\partial_t^2}{(x-y)^2H(y)}
+\partial_y(\frac{H(y)}{(x-y)^2})\partial_y
\nno\\&&+\partial_x(\frac{F(x)}{(x-y)^2})\partial_x
+\frac{\partial_{\si}^2}{(x-y)^2F(x)}]\phi.
 \ee
Under the decomposition of $\phi$ in some complete basis
\be \label{phi}
 \phi=\sum_{l,m}(x-y)\tilde{\phi}_{_{l
m}}(t,y_*^{})\chi_{_{lm}}(x,\si) ,
 \ee
and the tortoise-like coordinate defined by
\be \label{y*} dy_*^{}=\frac{dy}{H(y)}, \ee
the action reduces in the near horizon region ($y\rightarrow y_H$
and $H(y)\rightarrow 0$) to
\be \label{action2} S&=&-\frac{1}{2}\sum_{ll'mm'}\int
dtdxdy_*^{}d\si \tilde{\phi}_{_{l
m}}(t,y_*^{})\chi_{_{lm}}(x,\si)(-\pa_t^2+\pa_{y_*^{}}^2)\tilde{\phi}_{_{l'
m'}}(t,y_*^{})\chi_{_{l'm'}}(x,\si). \ee
After the integration over $x$ and $\si$ with the help of
\be \label{f} \int dx d\sigma \bar
\chi_{l'm'}^{}\chi_{lm}^{}=\delta_{ll'}\delta_{mm'},
 \ee
the action in the near horizon region reduces to an effective
2-dimensional one
\be \label{action3}
 S &=&-\frac{1}{2}\sum_{lm} \int dtdy_*^{}
\tilde{\phi}_{_{lm}}(t,y_*^{})(-\partial_t^2+\pa_{y_*^{}}^2)\tilde{\phi}_{_{lm}}(t,y_*^{}).
\ee
This indicates that the 2-dimensional effective metric near the
horizon is
\be \label{eadsc}
ds^2=-H(y)dt^2+\frac{dy^2}{H(y)}=-H(y)(-dt^2+dy_*^2). \ee
Therefore, following the discussion of section II, from
Eq.(\ref{N}) and Eq.(\ref{flux}), the flux which needed to cancel
the gravitational anomaly is
\be \label{phi1} \Phi=N^y_t|_{y_H^{}}=\frac {\pi} {12}  \left
(\frac{H'(y_H^{})}{4\pi}\right )^2 , \ee
so the Hawking temperature derived from the anomaly cancellation
method is $T=|H'(y_H^{})|/(4\pi)$.  Note that the Killing vector
which generates the bulk black hole horizon is
$\xi^a=(\partial_t)^a$.  A direct calculation shows the surface
gravity $\ka$ for the black hole horizon of the $AdS$
$\mathcal{C}$ metric is
\be \label{kac} \ka=\lim_{y\rightarrow
y_H^{}}\sqrt{\frac{(\xi^a\nabla_a \xi^c) (\xi^b \nabla_b
\xi_c)}{-\xi^d \xi_d}}=\frac{|H'(y_H^{})|}{2},
 \ee
which gives the horizon temperature by $T=\ka/(2\pi)$. Hence, the
Hawking temperature for the black hole horizon of the $AdS$
$\mathcal{C}$ metric obtained from Eq.(\ref{phi1}) is the same as
the one obtained from the standard way.

Recall that the black hole horizon in the  $AdS$ $\mathcal{C}$
metric is at\footnote{Another solution of $H(y)=y^2-2My^3=0$ is
$y_0=0$, which is sometimes called $AdS$ horizon and has zero
temperature from both Eq.(\ref{phi1}) and Eq.(\ref{kac}).}
$y_H^{}=1/(2M)$, its Hawking temperature is just
$T_H=|H'(y_H^{})|/(4\pi)=1/(8\pi M)=T_h$. This is a natural result
because the 3-dimensional Schwarzschild-like brane world black
hole described by Eq.(\ref{sbh}) is just the boundary of the black
hole in the bulk $AdS$ $\mathcal{C}$ spacetime \cite{emparan}. As
a result, the correct Hawking temperature for the 3-dimensional
Schwarzschild-like brane world black hole is reproduced by
applying the anomaly cancellation approach to the bulk $AdS$
$\mathcal{C}$ metric which generates it.


\section{Hawking Radiation of BTZ-like black hole on 2-brane}

Another subclass of the generalized ${\cal C}$ metric can be
expressed as \cite{cbook, ehm1}
\be \label{d}
ds^2=&&\frac{1+a^2x^2y^2}{(x-y)^2}[-\frac{H(y)}{(1+a^2x^2y^2)^2}(dt+ax^2d\si)^2
+\frac{dy^2}{H(y)}\nno\\&&+\frac{dx^2}{F(x)}+\frac{F(x)}{(1+a^2x^2y^2)^2}
(d\si-ay^2dt)^2] \ee
where
\be \label{hx1}
H(y)&&=-\ga-\La/6-2ly+k y^2-2My^3+(\ga-\La/6)a^2y^4,\nno\\
F(x)&&=\ga-\La/6+2lx-k x^2+2Mx^3-(\ga+\La/6) a^2x^4, \ee
with $a$ the rotation parameter, $\ga$ the acceleration parameter,
$l$ the NUT parameter and $k=1,0,-1$. When $a=0$, Eq.(\ref{d})
goes back to the $AdS$ $\mathcal{C}$ metric Eq.(\ref{adscm}). This
metric can be used to generate various kinds of 3-dimensional
brane world black holes. Similar to their four dimensional
counterparts, there are Hawking radiation and thermodynamics for
these 3-dimensional brane world black holes \cite{ehm1}. In the
following we take the BTZ-like brane world black hole as an
example {to show that the anomaly cancellation approach also
applies to these brane world black holes.} The discussions for
other cases are almost the same.

After making the coordinate transformations
\be \label{trans2}
 r=\frac{\sqrt{y^2+\la x^2}}{(x-y)} \quad {\rm
and} \quad \rho=\sqrt{\frac{1+kx^2-\tilde{a}^2x^2y^2/4}{y^2+\la
x^2}},
 \ee
and the choice of parameters $l=M=0$, $k=-1$, and $\ga-\La/6=1$,
the induced 3-dimensional metric on the RS 2-brane at $x=0$ is
then the BTZ-like brane world black hole \cite{ehm1,btz}
\be \label{bbtz}
 ds^2=-(k+\frac{\rho^2}{l_3^2}
+\frac{\tilde{a}^2}{4\rho^2})dt^2+(k+\frac{\rho^2}{l_3^2}
+\frac{\tilde{a}^2}{4\rho^2})^{-1}d\rho^2+\rho^2(d\si-\frac{\tilde{a}}{2\rho^2}dt)^2,
 \ee
where $\tilde{a}\equiv 2a$, which is the angular momentum, and
$\la\equiv \ga+\La/6\equiv -1/l_3^2$, which acts as the
cosmological constant on the RS 2-brane (at $x=0$).

The horizon of this brane world black hole is located at
$k+\rho^2/l_3^2 +\tilde{a}^2/(4\rho^2)=0$, and its Hawking
temperature is
\be \label{btzt} T_h=\frac{\rho_+^2-\rho_-^2}{2\pi\rho_+
l_3^2}=\frac{\sqrt{k^2+\la \tilde{a}^2}}{\pi \tilde{a}}\left
(\frac{-k-\sqrt{k^2+\la \tilde{a}^2}}{2}\right )^{1/2}
 \ee
with $\rho_\pm^2=(-k\pm\sqrt{k^2+\la \tilde{a}^2})/(-2\la)$.

Again, the horizon of this BTZ-like brane world black hole extends
into the bulk. And one can show that the bulk extension of the
brane world black hole horizon coincides with the black hole
horizon of the bulk generalized $\mathcal{C}$ metric (and that
both of them are minimal surfaces). Consequently, one can apply
the same procedure of the previous sections to check whether the
anomaly cancellation approach is applicable to the BTZ-like brane
world black hole.

Near the horizon, the action for  a massless scalar field $\phi$
in the generalized $\mathcal{C}$ metric becomes
\be \label{action4} S =&&-\frac{1}{2}\lim_{y\rightarrow y_H}\int
dtdxdy_*^{}d\si
\sqrt{-g}\phi\Box\phi \nno\\
=&& -\frac{1}{2}\sum_{ll'mm'}\int dtdxdy_*^{}d\si
\tilde{\phi}_{_{lm}}\psi_{_{lm}}[-(\pa_t+ay_H^2\pa_
\si)^2+\pa_{y_*^{}}^2]\tilde{\phi}_{_{l'm'}}\psi_{_{l'm'}},%
\ee
under the decomposition
$\phi=\sum_{lm}(x-y)\tilde{\phi}_{_{lm}}(\eta,
y_*^{})\psi_{_{lm}}(x,\si)$ and the tortoise coordinate $y_*^{}$.
The coefficient $ay_H^2$ corresponds to the horizon angular velocity
$\Omega_H$. One may also regard this term as  $U(1)$ gauge potential
$\mathcal{A}^t=ay_H^2$ like the treatment in \cite{iuw1, iuw2}. In
the near horizon region, one may make the coordinate transformation
\be \label{trans1} \varphi=\si-\Omega_H t \quad {\rm and} \quad
\eta=t, \ee
so that
\be \label{pt} \pa_\si=\pa_\varphi \quad {\rm and} \quad
\pa_t=-\Omega_H \pa_\varphi+\pa_\eta. \ee
Integrating the $x$ and $\varphi$ component of Eq.(\ref{action4}),
one gets the 2-dimensional effective action
\be \label{action5} S =&&-\frac{1}{2}\sum_{lm} \int d\eta dy_*^{}
\tilde{\phi}_{_l}(\eta, y_*^{})
(-\pa_{\eta}^2+\pa_{y_*^{}}^2)\tilde{\phi}_{_l}(\eta, y_*^{}).
 \ee
Eq.(\ref{action5}) indicates that the near horizon 2-dimensional
effective metric is of the same form as Eq.(\ref{eadsc}) and
requires the same flux as Eq.(\ref{phi1}) to cancel the
gravitational anomaly.  Obviously, the Killing vector that
generates the black hole horizon of generalized $\mathcal{C}$
metric is $\xi^a=(\pa_t)^a+ay_H^2(\pa_\si)^a$. Thus, the surface
gravity $\ka$ is again $|H'(y_H^{})|/2$. Therefore, the Hawking
temperature for the black hole horizon in the generalized
$\mathcal{C}$ metric is
\be \label{dt}
T_H=\frac{|H'(y_H^{})|}{4\pi}.
\ee

On the other hand, from Eq.(\ref{trans2}), on the 2-brane (i.e.
$x=0$), $\rho=1/y$ for ($y>0$). Meanwhile,
$H(y)=-\lambda+ky^2+a^2y^4=0$ gives $y_{\pm}^2=\frac{-k\pm
\sqrt{k^2+\la \tilde{a}^2}}{\tilde{a}^2/2}$, and the black hole
horizon is at $y=y_-$. Since one has the relation
$\rho_\pm=1/y_\mp$, it is easy to check that the Hawking
temperature of the black hole horizon in the bulk obtained from
Eq.(\ref{dt}) is
\be \label{bhtgc}
T_H=\frac{|H'(y)|}{4\pi}|_{y_-}=\frac{\sqrt{k^2+\la
\tilde{a}^2}}{\pi \tilde{a}}\sqrt{\frac{-k-\sqrt{k^2+\la
\tilde{a}^2}}{2}}.
 \ee
Clearly, one gets $T_H=T_h$, again. The reason is the same as
before, i.e., the BTZ-like brane world black hole described by
Eq.(\ref{bbtz}) is just the boundary of the black hole in the bulk
\cite{ehm1, emparan}. Therefore, through applying the anomaly
cancellation method to the bulk generalized $\mathcal{C}$ metric,
one can get the correct Hawking temperature of the BTZ-like brane
world black hole.


\section{Conclusions and Discussions}
In this letter, we showed that the anomaly cancellation method
proposed by Wilczek and his collaborators can be applied to the
3-dimensional brane world black holes generated by the generalized
$\mathcal{C}$ metrics in the RS scenario. Of course, one can apply
this method directly to the brane world black hole themselves.
However, the brane world black holes are different from the ordinary
black holes in general relativity. A significant difference is that
their horizons will extend into the extra dimensions, but not merely
on the brane, which means that the brane world black hole will
radiate both on the brane and in the bulk, e.g., \cite{Myers,
kanti2}. This property allows one to apply the anomaly cancellation
approach to the bulk spacetime to reproduce the Hawking temperature
of the brane world black holes. Although the 3-dimensional brane
world black holes and gravitational anomaly are only taken into
account here, they are easily generalized to the situations of
higher dimensional brane world black holes and with additional gauge
anomalies. In fact, the above calculations also hold for the
generalized ${\cal C}$ metrics (if one forgets the notion of brane
world black holes). Meanwhile, because this method does not involve
the dynamical equation, i.e., the Einstein equation, it also shows
that the Hawking radiation itself is not a dynamical effect but only
a kinematic one (notice that this method has been used in \cite{ks}
to get the Hawking radiation in acoustic black holes), which is in
accord with the studies about the acoustic black holes and the
uniformly accelerated reference frames \cite{unruh1, vi1, vi2, cg,
Sun}. The previously successful works of applying this approach to
various spacetime backgrounds suggest that there is a deep
relationship between the anomaly and Hawking radiation of the
horizon, and it is hopeful to extend this method to the more general
kinds of horizons. We shall consider such problems in a later work.

\begin{acknowledgments}\vskip -4mm
We thank Z. Chang and H.-Y. Guo for useful discussions. JRS would
like to thank R.-G. Cai, Y.-F. Cai and L.-M. Cao for helpful
discussions. This work is supported by  NSFC under Grant Nos.
90403023, 10605006, 10705048, 10775140 and Knowledge Innovation
Funds of CAS (KJCX3-SYW-S03).
\end{acknowledgments}


\begin{thebibliography}{07}

\bibitem{Hawking}
S.~Hawking, ``Black Hole Explosions?'', Nature (London) {\bf 248}
(1974) 30. ``Particle Creation by Black Holes'', Commun. Math.
Phys. {\bf 43} (1975) 199.

\bibitem{unruh}
W. G. Unruh,
\newblock ``Notes on black-hole evaporation,''
\newblock \PRD {\bf 14}, 870 (1976).

\bibitem{dr}
T.~Damoar and R.~Ruffini, ``Black-hole evaporation in the
Klein-Sauter-Heisenberg-Euler formalism'', \PRD {\bf 14} (1976)
332.

\bibitem{pw}
M.K.Parikh and F.~Wilczek, ``Hawking Radiation As Tunneling'',
Phys. Rev. Lett. {\bf 85} (2000) 5042.

\bibitem{cf}
S.~M.~Christensen and S.~A.~Fulling, ``Trace anomalies and the
Hawking effect'', Phys. Rev. {\bf D} {\bf 15} (1977) 2088.

\bibitem{rw}
S.~P.~Robinson and F.~Wilczek, ``Relationship between Hawking
Radiation and Gravitational Anomalies'', \PRL {\bf 95} (2005)
011303.

\bibitem{iuw1}
S.~Iso, H.~Umetsu and F.~Wilczek, ``Hawking Radiation from Charged
Black Holes via Gauge and Gravitational Anomalies'',  \PRL {\bf
96} (2006) 151302.

\bibitem{iuw2}
S.~Iso, H.~Umetsu and F.~Wilczek, ``Anomalies, Hawking Radiation
and Regularity in Rotating Black Holes'', \PRD {\bf74} (2006)
044017.

\bibitem{witten} L.~Alvarez-Gaume and E.~Witten,
``Gravitational Anomalies,'' \NPB234, 269 (1984).

\bibitem{ms}
K.~Murata and J.~Soda, ''Hawking radiation from rotating black
holes and gravitational anomalies'', Phys. Rev. {\bf D} {\bf74}
(2006) 044018.

\bibitem{vd}
E. C. Vagenas and S. Das, ``Gravitational anomalies, Hawking
radiation, and spherically symmetric black holes'', JHEP 10 (2006)
025.

\bibitem{se}
M.~R.~Setare, ``Gauge and gravitational anomalies and Hawking
radiation of rotating BTZ black holes'', Eur. Phys. J. {\bf C}
{\bf 49} (2007) 865.

\bibitem{xc}
Z.~Xu and B.~Chen, ``Hawking radiation from general Kerr-(anti)de
Sitter black holes'', Phys. Rev. {\bf D} {\bf75} (2007) 024041.

\bibitem{jw}
Q.~Q.~Jiang, S.~Q.~Wu, ``Hawking radiation from rotating black
holes in anti-de Sitter spaces via gauge and gravitational
anomalies'', Phys. Lett. {\bf B} {\bf647} (2007) 200.

\bibitem{jwc}
Q.~Q.~Jiang, S.~Q.~Wu and X.~Cai, ``Hawking radiation from the
dilatonic black holes via anomalies'', Phys. Rev. {\bf D}{\bf 75}
(2007) 064029.

\bibitem{xlz}
Kui Xiao, Wenbiao Liu, Hongbao Zhang, ``Anomalies of the
Achucarro¨COrtiz black hole '', Phys. Lett. {\bf B} {\bf647}
(2007) 482-485.

\bibitem{sk}
H. Shin, W. Kim, ``Hawking radiation from non-extremal D1-D5 black
hole via anomalies'', JHEP 06 (2007) 012.

\bibitem{drv}
S. Das, S. P. Robinson and E. C. Vagenas, ``Gravitational anomalies:
A Recipe for Hawking radiation", arXiv:0705.2233.

\bibitem{ch}
B.~Chen and W.~He, ``Hawking Radiation of Black Rings from
Anomalies", arXiv:0705.2984

\bibitem{mm}
K. Murata, U. Miyamoto, ``Hawking radiation of a vector field and
gravitational anomalies'', arXiv:0707.0168 [hep-th]

\bibitem{bk}
R. Banerjee and S. Kulkarni, ``Hawking Radiation and Covariant
Anomalies" arXiv:0707.2449[hep-th]; ``Hawking Radiation, Effective
Actions and Covariant Boundary Conditions", arXiv:0709.3916[hep-th].

\bibitem{ks}
W. Kim, H. Shin, ``Anomaly analysis of Hawking radiation from
acoustic black hole'', JHEP 07 (2007) 070.

\bibitem{RS}
L. Randall, R. Sundrum, ``An Alternative to Compactification'',
\PRL {\bf 83}, 23 (1999).

\bibitem{Chr} A. Chamblin, S. Hawking and H Reall, ``Brane-world
black holes'' \PRD {\bf 61}, 065007 (2000).

\bibitem{ehm1}
R. Emparan, G. Horowitz, R. Myers, ``Exact description of black
holes on branes'', JHEP01, 007 (2000); ``Exact description of
black holes on branes II: comparison with BTZ black holes and
black strings'', JHEP01, 021 (2000).

\omits{\bibitem{dmpr} N. Dadhich, R. Maartens, P. Papadopoulos, V.
Rezania, ``Black holes on the brane'', \PLB {\bf 487}, 1 (2000).}

\bibitem{kanti}
P. Kanti, K. Tamvakis, ``Quest for localized 4D black holes in
brane worlds'' \PRD {\bf 65}, 084010 (2002).

\bibitem{cfm}
R. Casadio, A. Fabbri and L. Mazzacurati, ``New black holes in the
brane world?'' \PRD {\bf 65}, 084040 (2002).

\bibitem{aliev}
A. N. Aliev, ``Charged rotating black holes on a 3-brane", Phys.
Rev. D71, 104027 (2005).

\bibitem{emparan}
R. Emparan, ``Black hole entropy as entanglement entropy: a
holographic derivation'', JHEP06, 012 (2006).

\bibitem{unruh1}
W. G. Unruh, ``Experimental black-hole evaporation?'', \PRL {\bf
46}, 1351 (1981).

\bibitem{vi1}
M. Visser, ``Acoustic black holes: Horizons, ergospheres, and
Hawking radiation'', \CQG {\bf 15}, 1767 (1998).

\bibitem{vi2}
M. Visser, ``Hawking radiation without black hole entropy'', \PRL
{\bf 80}, 3436 (1998).

\bibitem{cg}
C.-G. Huang and H.-Y. Guo, ``A new kind of uniformly accelerated
reference frame", \IJMPD {\bf 15}, 1035 (2006).

\bibitem{Sun}
C.-G. Huang and J.-R.~Sun, ``Thermodynamic Properties of
Spherically-Symmetric, Uniformly-Accelerated Reference Frames'',
gr-qc/0701078, to appear in \CTP.

\bibitem{bert1}
R.~Bertlmann, {\it Anomalies In Quantum Field Theory}
(Oxford Science Publications, Oxford, 2000).

\bibitem{bert2}
R.~Bertlmann and E.~Kohlprath, ``Two-dimensional gravitational
anomalies, Schwinger terms and dispersion relations,'' Ann.\
Phys.\ (N.Y.) {\bf 288}, 137 (2001). [arXiv:hep-th/0011067].

\bibitem{cbook}
H. Stephani, D. Kramer, M. Maccallum, C.Hoenselaers and E. Herlt,
``Exact Solutions of Einstein's Field Equations'', Cambridge
University Press, Second editon.

\bibitem{cmetric}
J. F. Plebanski and M. Demianski, ``Rotating, charged, and
uniformly accelerating mass in general relativity'', Ann. Phys.
(NY) {\bf 98}, 98 (1976).

\bibitem{btz} M. Ba{\~n}ados, C. Teitelboim and J. Zanelli, ``The
black hole in three-dimensional spacetime'', \PRL {\bf 69} (1992)
1849.

\bibitem{Myers}
R. Emparan, G. T. Horowitz and R. C. Myers, ``Black Holes Radiate
Mainly on the Brane'', \PRL {\bf 85}, 499 (2000).

\bibitem{kanti2}
P. Kanti, ``Black holes in theories with large extra dimensions: a
review'', \IJMPA, {\bf 19}, 4899 (2004).



\end{thebibliography}
\end{document}